\documentclass[12pt]{iopart}
\usepackage{epsfig}
\usepackage{graphicx}

\begin{document}

\title{Jet shower evolution in medium and dijet asymmetry in Pb+Pb collisions at the LHC}

\author{Guang-You Qin}

\address{Department of Physics, Duke University, Durham, North Carolina, 27708, USA}
\ead{qin@phy.duke.edu}

\begin{abstract}
We study the evolution of a partonic jet shower propagating through a quark-gluon plasma. 
Combining the in-medium evolutions of  the leading parton and shower gluons, we compute the depletion of the energy from the jet cone by 
dissipation through elastic collisions with medium constituents, by scattering of shower partons to larger angles, and by radiation outside the jet cone. 
Numerical results are presented for the nuclear modification of dijet asymmetry in Pb+Pb collisions at the LHC.
\end{abstract}

%\submitto{\JPA} % Comment out if separate title page not required

%\maketitle

\section{Introduction}

Jet quenching has been studied in some details (e.g., see \cite{Armesto:2011ht, Bass:2008rv}) in relativisitic heavy-ion collisions due to the fact that hard
partons interact with the traversed mater and lose some of the initial energy by elastic and inelastic collisions with medium constituents, providing powerful
probes to the hot and dense matter produced in these reactions \cite{Majumder:2010qh, Bjorken:1982tu, Zakharov:1996fv, Baier:1996kr}. 
One of the most spectacular jet measurements from the first heavy-ion runs at the CERN large Hardon Collider (LHC) is the nuclear modification of the energy asymmetry between two correlated jets
emitted in opposite azimuthal direction around the beam axis in Pb+Pb collisions at $\sqrt{s_{NN}}=2.76$~TeV \cite{Aad:2010bu, Chatrchyan:2011sx}. 

In this work, we study the medium modification of a parton shower propagating through a quark-gluon plasma \cite{Qin:2010mn}. 
As the jet shower traverses the medium, we calculate the energy loss experienced  by the jet shower within a cone angle defined as 
$R=\sqrt{(\Delta \phi)^2 + (\Delta \eta)^2}$, which includes  
collisional energy loss by  the leading parton and shower gluons through elastic
collisions with medium constituents, as well as the energies carried by shower gluons that are radiated or scattered outside the jet cone. 
We present the numerical results for the nuclear modification of dijet energy asymmetry in Pb+Pb collisions at the LHC.

\section{Jet shower evolution in medium}

In our model, a jet shower is described by two quantities: the energy $E_L$ of  the leading parton and the double-differential distribution $f_g(\omega,
k_\perp^2,t) = dN_g(\omega, k_\perp^2, t)/d\omega dk_\perp^2$ of shower gluons, where $\omega$ denotes the gluon energy and $k_\perp$ the transverse
momentum with respect to the jet axis.
For a jet defined by a cone angle $R$, the energy contained in the jet cone is given by
\begin{eqnarray}
\label{eq:Eg}
E_{J}(R) = E_L + E_g(R) = E_L + \int_{R} \omega d\omega dk_\perp^2 f_g(\omega, k_\perp^2)  , \ \ \ 
\end{eqnarray}
where the subscript $R$ denotes the integration taken within the jet cone, $k_\perp<\omega R$.

The evolution of a jet shower propagating through medium is governed by two equations, i.e., the evolutions of the leading parton energy and shower gluon distribution:
\begin{eqnarray}
\label{eq:EL}
E_{\rm L}(t_f) = E_{\rm L}(t_i) - \int \hat{e}_L(t) dt  - \int \omega d\omega dk_\perp^2 dt\, \frac{dN_g^{\rm med}}{d\omega dk_\perp^2 dt}  ,
\end{eqnarray}
\begin{equation} 
\label{eq:dG/dt}
\frac{d}{dt}f_g(\omega, k_\perp^2, t) = \hat{e} \frac{\partial f_g}{\partial \omega}
  + \frac{1}{4} \hat{q} {\nabla_{k_\perp}^2 f_g} +  \frac{dN_g^{\rm med}}{d\omega dk_\perp^2 dt} .
\end{equation}
Here we include both the contributions from elastic collisions and medium-induced radiation, as well as that from transverse momentum broadening experienced by shower gluons. 
Note that a source term from vacuum radiation should be also included for the interference with medium-induced radiation. In this application, we assume
that jets experiences vacuum radiation before interacting with medium, thus vacuum radiation serves as the initial conditions for jet shower in-medium evolution. 

After solving the above jet shower evolution equations, one may calculate the total energy loss from the jet cone, 
\begin{eqnarray}
\label{eq:DeltaE}
\Delta E_{J} = E_{J}(t_i, R) - E_{J}(t_f, R) .
\end{eqnarray}

A few inputs must be supplemented to solve the above jet shower evolution equations and  calculate the energy loss from the jet cone. 
The initial gluon distribution, $f_g(\omega, k_\perp^2,t_i)$, before interacting with the thermal medium at time $t_i$, is generated from PYTHIA
\cite{Sjostrand:2007gs}. 
We impose a formation time cut-off, i.e., only gluons with a formation time $\tau_f = 2Ex(1-x)/k_\perp^2$ smaller than $t_i$ are radiated, 
where $x=\omega/E$ denotes the energy fraction of the radiated gluons. 
The rate for the medium-induced gluon radiation is taken from the higher-twist formalism of jet quenching \cite{Wang:2001ifa,
Majumder:2009ge}:
\begin{eqnarray}
\label{eq:Gmed}
\frac{dN_g^{\rm med}}{d\omega dk_\perp^2 dt} = \frac{2\alpha_s}{\pi} \frac{x P(x) \hat{q}(t)}{\omega k_\perp^4}   \sin^2 \frac{t - t_i}{2\tau_f} ,
\end{eqnarray}
where $P(x)$ is the vacuum splitting function.  
We further relate two transport coefficients by the fluctuation-dissipation theorem, $\hat{q} = 4T\hat{e}$, assuming that the medium is in thermal equilibrium.
When solving for the radiated gluon distribution $f_g(\omega, k_\perp^2, t)$, a lower cut-off of $2$~GeV is imposed on the energy of shower gluons. 

\section{Nuclear modification of dijet asymmetry at the LHC}

We now calculate the medium modification of the dijet asymmetry in Pb+Pb collisions at the LHC. 
The space-time profile of the medium temperature $T(\vec{r}_\perp, t)$ at midrapidity is modelled as follows.
The initial entropy density $s\sim T^3$ of the medium is set to be proportional to the density of participating nucleons in the colliding nuclei, with 
the nuclear density distributions taken as Woods-Saxon profiles.
The medium created in Pb+Pb collisions at $\sqrt{s_{NN}}=2.76$~TeV is assumed to thermalize at $t_0 = 0.6$~fm/$c$, at which the temperature of the hottest
point in central collisions is set to be $T_0 = 520$~MeV.
The time evolution of medium is modeled by a one-dimensional boost-invariant expansion, {\em i.e.}, the temperature falls with time as $t^{-1/3}$. 
For jet shower evolution, we assume that it experiences vacuum radiation before $t_0$, after which jet-medium interaction starts until the local temperature of the medium drops
below $T_c=160$~MeV. 
The transport coefficients are scaled according to the temperature of the medium, $\hat{q} \propto T^3$, with a constant factor adjusted to fit the experimental
data. 

\begin{figure}[htb]
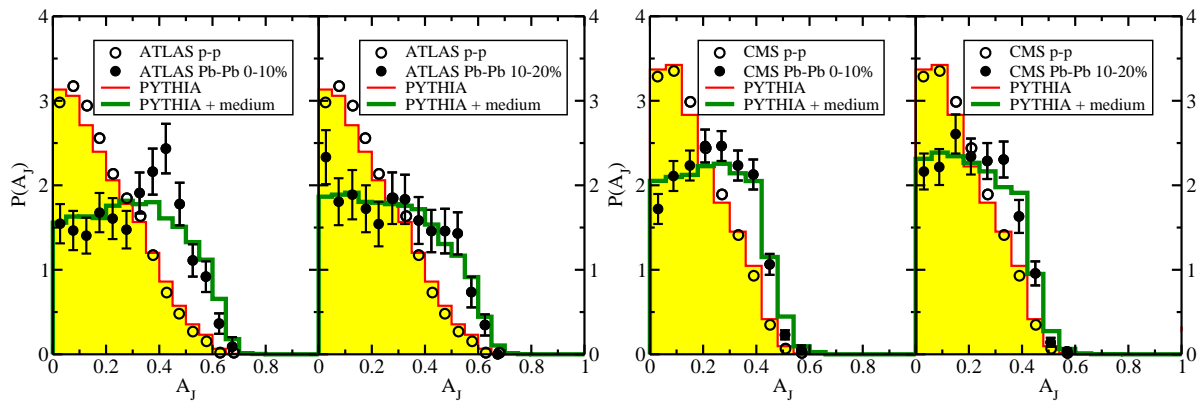

\includegraphics[width=0.5\linewidth]{dijet_atlas.eps} 
\includegraphics[width=0.5\linewidth]{dijet_cms.eps} 
\caption{(Color online) 
Distribution of dijet energy asymmetry factor $A_J$ for $p+p$ and Pb+Pb collisions at $\sqrt{s_{NN}}=2.76$~TeV at the LHC. Left two panels: $0$-$10$\%
and $10$-$20$\% centralities compared to the ATLAS \cite{Aad:2010bu}. Right two panels: $0$-$10$\% and $10$-$20$\% centralities compared to CMS \cite{Chatrchyan:2011sx}. }
\label{dijet_Aj}
\end{figure}

The results of the medium-modification of dijet asymmetry factor $A_J$ is shown in Fig. \ref{dijet_Aj}, where $A_J$ is defined as
\begin{equation}
\label{eq:AJ}
A_J = \frac{E_{T,1} - E_{T,2}}{E_{T,1} + E_{T,2}} ,
\end{equation}
with $E_{T,i}, (i=1,2)$ the transverse energy of the leading and sub-leading jet, respectively.
The ATLAS Collaboration measured this quantity with the trigger jet $E_{T,1}>100$~GeV and the second jet in the opposite hemisphere $\Delta\phi>\pi/2$
with $E_{T,2}>25$~GeV. 
For CMS Collaboration, these values are $E_{T,1}>120$~GeV, $\Delta\phi>2\pi/3$, and $E_{T,2}>25$~GeV. 

The vacuum dijet events for p+p collisions at the LHC energies shown by the shaded yellow mountains are generated from PYTHIA \cite{Sjostrand:2007gs}, where jets are reconstructed using 
anti-$k_T$ algorithms \cite{Cacciari:2008gp}. Note a Gaussian smearing with width $\propto \sqrt{E_{J}}$
is applied  here to taken into account the detector response and other smearings; this reduces some amount of very symmetric dijet events.  
The modification of each dijet event in Pb+Pb collisions is obtained as follows.
For each dijet event, its production point is sampled according to the distribution of the binary nucleon-nucleon collisions from colliding two lead nuclei.
Then additional energy loss from the jet cone due to jet-medium interaction is applied and the dijet asymmetry $A_J$ distribution is recalculated from the surviving dijet events as shown by the
thick green lines. 
The trigger bias is approximated by letting the leading jet propagate along the shorter path, and the subleading jet
propagate along the other direction when dijets are asymmetric ($A_J>0.1$). 
For nearly symmetric jet pairs ($A_J<0.1$), we do not apply such trigger bias treatment.

As expected, the energy asymmetry of dijes is significantly increased by the in-medium evolution and more prominent in the most central Pb+Pb collisions than in less central events. 
By fitting to the data we obtain the length-averaged transport coefficient in central collisions $\langle \hat{q} \rangle \approx 0.9$~GeV$^2$/fm (with about $20$\% difference between ATLAS and CMS for the best descriptions of the data). 
This corresponds to $\hat{q} \approx 2$~GeV$^2$/fm at $T=400$~MeV, the highest temperature available in Au+Au collisions at Relativistic Heavy Ion Collider (RHIC).

\section{Summary}

In summary, we have presented a simplified but realistic model for studying the evolution of a jet shower propagating in a quark-gluon plasma.
We have applied the model to compute the nuclear modification of dijet energy asymmetry in Pb+Pb collisions at the LHC. 
The observed dijet asymmetry can be described with values of parton transport coefficients similar to 
those describing jet quenching data at RHIC.  
Some further directions include a complete Monte-Carlo simulation of jet shower evolution in medium and reconstruction, and 
realistic event-by-event hydrodynamical simulation of fluctuating background (e.g., see \cite{Qin:2010pf}) and its effect on dijet energy asymmetry \cite{Cacciari:2011tm}.

\section*{Acknowledgments}

G.-Y.~Q. thanks B. M\"uller for helpful discussions. This work was supported in part by Grants No. DE-FG02-05ER41367 and No. DE-SC0005396 from the U.S. Department of Energy.

\section*{References}

%\begin{thebibliography}{10}

\bibliographystyle{h-physrev5.bst}
\bibliography{GYQ_refs.bib}

%\end{thebibliography}

\end{document}